\documentclass[preprint,12pt]{aastex}

\newcommand{\degree}{\mbox{$^{\circ}$}}

\newcommand{\as}{\mbox{\arcsec}}

\def\lsim {$\rlap{\raise.4ex\hbox{$<$}}\lower.55ex\hbox{$\sim$}\,$}

\newcommand{\msun}{\mbox{M$_\odot$}}

\newcommand{\mean}[1]{\mbox{$\langle#1\rangle$}} 
 
\input{epsf}

\newcommand{\tie}{\mbox{$T^{env}_{iso}$}}

\begin{document}

               
\title {MUSTANG 3.3 Millimeter Continuum Observations of Class 0 Protostars}
\author {Yancy L. Shirley\altaffilmark{1}, \altaffilmark{2},
	 Brian S. Mason\altaffilmark{4},
         Jeffrey G. Mangum\altaffilmark{4}, 
	 David E. Bolin\altaffilmark{1} \altaffilmark{3},
	 Mark J. Devlin\altaffilmark{5},
	 Simon R. Dicker\altaffilmark{5}, and
	 Phillip M. Korngut\altaffilmark{5}
} 

\altaffiltext{1}{Steward Observatory, University of Arizona, 933 Cherry Ave., Tucson, AZ 85721}
\altaffiltext{2}{Ajunct Astronomer, National Radio Astronomy Observatory}
\altaffiltext{3}{UA/NASA Space Grant Undergraduate Research Internship Program, 2008-2009}
\altaffiltext{4}{NRAO, 520 Edgemont Road, Charlottesville, VA 22903-2475}
\altaffiltext{5}{Unversity of Pennsylvania, 209 South 33rd Street, Philadelphia, PA 19104-6396}
 
\begin{abstract}
 
We present observations of six Class 0 protostars at $3.3$ mm ($90$ GHz) 
using the 64-pixel MUSTANG bolometer camera on the 100-m Green Bank Telescope.
The $3.3$ mm photometry is analyzed along with shorter wavelength observations
to derive spectral indices ($S_{\nu} \propto \nu^{\alpha}$) of the measured emission.
We utilize previously published dust continuum radiative transfer models to estimate
the characteristic dust temperature within the central beam of our observations.
We present constraints on the millimeter dust opacity index $\beta$
between $0.862$ mm, $1.25$ mm, and $3.3$ mm.  $\beta_{mm}$ 
typically ranges from $1.0$ to $2.4$ for Class 0 sources.  The relative contributions
from disk emission and envelope emission are estimated at 3.3 mm.  
L483 is found to have negligible disk emission at 3.3 mm while L1527 is dominated
by disk emission within the central beam.  The $\beta_{mm}^{disk} \leq 0.8 - 1.4$
for L1527 indicates that grain growth is likely occurring in the disk.
The photometry presented in this paper may be combined with
future interferometric observations of Class 0 envelopes and disks.
\end{abstract}



\section{Introduction}

One of the fundamental problems in studies of low-mass ($M <$ few \msun )
star formation is to track the flow of material from the dense molecular
cloud core through the disk and onto the protostar.    Several theoretical
models exist that describe the hydrodynamic collapse of protostellar
envelopes (e.g., Larson 1969; Penston 1969; Hunter 1977; Shu 1977; Tereby, Shu,
\& Cassen 1984; Whitworth \& Summers 1985; Foster \& Chevalier 1993;
Galli et al. 1993; Li \& Shu 1996; McLaughlin \& Pudritz 1997;
Fatuzzo et al. 2004; Adams \& Shu 2007).  Observations of the density, temperature,
and kinematic structure of the dense cores and disk can distinguish
between these theoretical models.

Since the timescale
for formation of a solar mass star is typically longer than a million years,
we must track the evolution of material by observing protostellar objects
in different evolutionary states and piece together
an evolutionary theory for the collapse of dense cores and the
subsequent growth and dissipation of a protostellar disk 
(e.g, Shu, Adams, \& Lizano 1987; Andr\'e et al. 1993, 2000).
The Class 0 phase is of particular interest because it represents the earliest phase 
of star formation after the formation of a first hydrostatic core.
Class 0 objects are defined as protostars that have yet to accrete the 
majority of their final mass (Andr\'e et al. 2000) although in practice this
definition is difficult to confirm for individual objects.  
Observationally, Class 0 protostars are defined
as protostars which emit more than $0.5$\% of their luminosity at wavlengths
longer than $350$ \micron\ (see Andr\'e  et al. 1993).
They are deeply-embedded in a
dense gas and dust envelope (Shirley et al. 2000), they typically drive strong 
molecular outflows (Bontemps et al. 1996), 
and their luminosities are dominated by a variable accretion luminosity (Evans et al. 2009).  
Over the past decade, deep mid-infrared surveys have cataloged the population
of Class 0 protostars in nearby molecular clouds (Dunham et al. 2008,
Evans et al. 2009).

Observations at submillimeter and millimeter wavelengths of Class 0
protostars are useful for constraining the density and
temperature structure of envelopes and disks.  The emission is
dominated by optically thin dust continuum emission (Adams 1991, Shirley et al. 2003).  
Radiative transfer modeling of the submillimeter continuum has determined
the envelope density structure for several Class 0 protostars
(e.g., Shirley et al. 2002, J\o rgensen et al. 2002) which 
is a particularly good discriminant between theoretical
models of inflow and collapse in the protostellar envelope
(see Myers et al. 2000; Andr\'e et al. 2000).  The majority of the modeling
has been focused on single-dish (sub)millimeter observations with modest ($10$\as )
resolution; but, in order to study
the connection between the envelope and the disk and in order to study
the properties of Class 0 disks themselves, (sub)millimeter interferometric
observations are needed; however, the interferometric observations must be
modeled in combination with single-dish observations (which provide the 
zero spacing data)
to constrain the amplitude of model visibilities.  Furthermore, combined
interferometric plus single-dish observations are needed at multiple
wavelengths to constrain the properties of the dust emission
(i.e. $\beta$, the dust opacity index) which is a large source
of uncertainty in the state-of-the-art radiative transfer models.

The spectral energy distribution (SED) of several famous Class 0 sources
have been observed from mid-infrared through centimeter wavelengths; however, 
there is a gap in the observations of an order of magnitude in wavelength 
in the millimeter spectrum.  Traditionally, the longest wavelength 
bolometer arrays utilized for star formation studies 
operate in the 1 mm band (e.g. MAMBO 1.25 mm, 
SIMBA 1.2 mm, BOLOCAM 1.1mm, AzTEC 1.1 mm).
Recently, a new bolometer camera, MUSTANG, that operates at 3.3 mm (90 GHz) was
built for the 100-m Green Bank Telescope\footnote{The Green Bank Telescope
is operated by the National Radio Astronomy Observatory.  The National Radio Astronomy 
Observatory is a facility of the National Science Foundation operated under cooperative
agreement by Associated Universities, Inc.}
This region of the spectrum is of
interest because observations at longer wavelengths provide a larger lever-arm
for calculating dust opacity properties.  At these long wavelengths, the
emission from a protostellar disk may become an
important component to the total flux.  MUSTANG observations can provide
the zero spacing data for future interferometric observations (e.g. with ALMA) 
of these sources to study the envelope and disk emission at a longer millimeter 
wavelength than previously possible.

In this study, we have observed six Class 0 protostars with MUSTANG at 3.3 mm. 
We describe the reduction techniques and present calibrated photometry
for the sources (\S2). 
We present an analysis
of the spectral indices between the observed bands and published fluxes short-ward
of our 3.3 mm observations.  We constrain the properties
of the dust emission (\S3), accounting for emission from both the
envelope (\S3.1) and disk (\S3.2).

\section{MUSTANG Observations and Image Reduction}

We observed six Class 0 protostars (see Table 1) with the MUSTANG 3.3 mm camera
(Dicker et al. 2006, 2008) using the 100-m Green Bank
Telescope (GBT, Jewell \& Prestage 2004).  MUSTANG is a 64-pixel bolometer camera with
an 18 GHz wide continuum filter centered at 90 GHz (3.3 m).  
The pixels are transition edge-sensors spaced at
$\theta_{FWHM}/2\sim 4''$ with an instantaneous field-of-view of
$40$\as .  The theoretical beamsize is $8.5$\as\ on the GBT , however,
the illumination pattern plus surface inaccuracies result in slightly
larger beamsizes of $\sim 11''$ which we characterize with frequent
measurements of bright secondary calibrator sources in each observing
run.  Observations were made over three days (February 6, April 24,
and May 8) in 2009.  At the start of each run, observations of a bright
compact source (e.g. quasars) are used to solve for primary aperture wavefront phase
errors using the ``Out of Focus'' (OOF) holography technique
(Nikolic et al. 2007; Schwab \& Hunter, in preparation). 
The solutions were applied to the active GBT surface.
After the surface is calibrated, we observed the protostars using a
centrally-weighted daisy scanning pattern. This scan pattern strikes a
balance between modulating the sky quickly to beat-down detector noise
drifts, and accumulating integration time on a single, small region of
interest ($\sim 1'$ in diameter). Typical scan speeds on the sky are
$\sim 30$~-~$45''/s$.

Images are reduced by IDL routines through a custom pipeline.
Detector gains are determined by pulsing an internal calibration
lamp periodically.  A ``common mode'' template is computed as a function of time by
  taking the median of the set consisting of each good detector sample
  for a given integration.  This common mode template can optionally be low-pass filtered.
The common mode template is fit to and subtracted from each
  detector timestream. Most systematic effects, such as atmospheric
  emission fluctuations, make almost identical contributions to each
  detectors' data. The common mode is highly effective at removing
  these systematics. It also, however, filters out astronomical
  information on spatial scales larger than the instantaneous camera
  field-of-view ($40'' \times 40''$). The optional low-pass filtering mitigates
  this effect.
A low-order polynomial is fit to and subtracted from each detector timestream
  individually. The order of the polynomial is chosen based on the
  duration of the scan in question to have 1 degree of freedom per $\sim 10$
  seconds of data, depending on the stability of the data in question.
Individual detector weights are assigned based on the variance
  of the fully cleaned detector timestream. A preliminary SNR map is
  made before this calculation using the detector white noise level as
  an initial weight estimate; segments of the timestream corresponding
  to regions of the map with SNR higher than 5 are excluded from the
  detector weight calculation, to avoid biasing the weights by the
  presence of bright signal.
The data are corrected for atmospheric opacity with the opacity values for each observing 
session calculated from
publicly available National Weather Service data\footnote{{\tt
    http://www.gb.nrao.edu/$\sim$rmaddale/Weather/index.html}}.
Finally, the data are gridded into a map.

This procedure is iterated once to back out the effect of sky domain
signal on the common mode and polynomial solutions. At the signal
levels in our maps, further iteration does not have a significant
effect on the results.  Further details on the data analysis procedures
can be found in Dicker et al. (2009) and Mason et al. (2010).
The MUSTANG 3.3 mm images are shown in Figure 1.

The common mode subtraction used to remove atmospheric emission 
also has the effect of removing some large angular-scale source signals. 
We have characterized this effect with Monte Carlo simulations using the calculated
model envelope surface brightness profile of B335 from Shirley et al. (2010), 
with the range of data reduction parameters used in our actual image production scripts.  
These simulations are passed through our full data reduction pipeline so their results 
incorporate the effects of all stages of the data reduction process.  Of particular significance 
is the cutoff frequency of the low-pass filter applied to the common mode template 
(typically 0.1 Hz - 0.2 Hz for the data we present). We find that for B335, the peak surface 
brightness of the filtered envelope is 80\% to 87\% of the envelope as it would be seen by the GBT+MUSTANG 
in the absence of common mode subtraction.  By way of comparison, the map obtained with a pure common 
mode subtraction ({\it i.e.}, no low-pass filtering of the common mode template) retrieves only 
71\% of the peak central surface brightness.  Note that this analysis includes the effects of the
error beam of the telescope (discussed in Mason et al. 2010). 
The maximum effect, at the center of the envelope, causes $\sim 11$\% more power to be coupled 
in from the extended emission than would be seen in the absence of the error beam.  
Since we do not know, a priori, the disk contribution at 3.3 mm (\S3.2), we have only
characterized the surface brightness recovery of the envelope.

Observations were made of standard calibrators Ceres (February 6),
Neptune (April 24), and CRL2688 (May 8).  The final maps were calibrated in
mJy/beam and in mJy in a $20$\as\ aperture by comparing the peak flux
and flux in a $20$\as\ aperture respectively for the observed calibrators.  We limit the
aperture photometry to apertures $\leq 1/2$ the field-of-view of the
MUSTANG array.  The total predicted flux at 3.3 mm from Ceres was 465 mJy on
February 6 (Thomas Mueller, private communication).  A brightness
temperature of $142 \pm 12$ K was assumed for Neptune observations
(Weiland et al. 2010) in April.  The total flux from CRL2688 was
bootstrapped from previous MUSTANG observations over the 2009 season
and was calculated to be 141 mJy.  Aperture photometry on the Class 0
sources were performed to determine the peak voltage and the total
voltage in a $20$\as\ aperture which was then multiplied by the
appropriate calibration factor (mJy/Volts) for each day (Table 1). We note that
the weather was dry with low winds on February 6, but that the weather
was much wetter on April 24 and May 8. Calculated zenith opacities
varied from $\tau_{3.3} = 0.074$ to $0.25$.  Taking into account the fact that
errors in the assumed zenith opacity also affect the celestial
calibrator sources, we estimate that even in the most extreme case the
overall uncertainty in our results due to the uncertainty in the
zenith opacity is $< 5\%$. This is substantially less than the $8\%$ -
$10\%$ uncertainty in Neptune and Ceres absolute temperature scale.
We therefore assume a $10$\% systematic flux calibration uncertainty at 3.3 mm.

\section{Results}

In this section we constrain the dust opacity index ($\beta$; defined such that
$\kappa_{\nu} \propto \nu^{\beta}$) at millimeter wavelengths using estimates
for the dust temperature calculated from dust continuum radiative
transfer models.  We first
calculate the spectral index at (sub)millimeter wavelengths, then develop
a method for calculating the appropriate single dust temperature to characterize
emission within a central aperture.  We use the calculated dust
temperatures to determine the range in $\beta$.  Finally, in \S3.2 we estimate
the relative contributions to the observed $\beta$ from the envelope and disk
for each of the Class 0 source in our sample.

\subsection{Analyzing Spectral Indices}

We characterize the emission between two wavelengths by assuming that
the SED follows a power-law ($S_{\nu} \propto \nu^{\alpha}$) and calculating
the spectral index
\begin{equation}
\alpha_{\lambda_1/\lambda_2} = \frac{\log(S_{\lambda_1}/S_{\lambda_2})}{\log(\lambda_2/\lambda_1)} \;\; .
\end{equation}
The spectral index is an empirical property of the observed SED that
can be related to the underlying properties of the emission.

We have tabulated the observed spectral indices from 0.86 mm to 3.3 mm 
in Table 1.  The spectral index between 1.25 mm and 3.3 mm is calculated
from the peak flux density (mJy/beam) in both the MAMBO 1.25 mm observations
of M\"otte \& Andr\'e (2001) and our MUSTANG observations.  A direct comparison
is feasible since the solid angles of the IRAM 30-m at 1.25 mm and the GBT 100-m
at 3.3 mm are both approximately equivalent to the solid angle of a
Gaussian beam with FWHM of $11$\as .  Unfortunately, we were
not able to make the same peak
flux density comparison with SCUBA observations at 0.862 mm since
the effective beamsize of the JCMT is $16$\as .  Therefore, we calculate the
spectral index between 0.86 mm and 3.3 mm using matched $20$\as\ diameter aperture
photometry.  The typical millimeter spectral index varies between $\alpha_{1.25/3.3} 
= 2.6$ to $3.8$ with an average of $\mean{\alpha_{1.25/3.3}} = 3.2$ for this sample.
Due to L1448NW being at the edge of the map, we were unable to calculate
the flux in a $20$\as\ aperture at 3.3 mm for this source.  Excluding L1448NW, the average
$\mean{\alpha_{0.86/3.3}} = 3.0$ is very similar to the spectral index calculated
at 1.25 mm for the same sources.

If we assume the dust opacity ($\kappa_{\nu}$ cm$^2$ gram$^{-1}$) 
follows a single-power law at 
(sub)millimeter wavelengths ($\kappa_{\nu} \propto \nu^{\beta}$), 
then the dust opacity index, $\beta$, may be found from the ratio of fluxes
at the two wavelengths
\begin{equation}
\frac{S_{\lambda_1}}{S_{\lambda_2}} = \left( \frac{\lambda_2}{\lambda_1} \right)^{(3+\beta)}
\frac{exp(h\nu_2/kT_{d}) - 1}{exp(h\nu_1/kT_{d}) - 1} \;\;\;.
\end{equation}
Equation 2 assumes that a single dust temperature, $T_d$, characterizes the emission at 
both wavelengths
The derived $\beta$ can
be sensitive to the choice of the dust temperature.  For instance, at submillimeter
wavelengths of $442$ and $862$ \micron , $\beta$ varies by a factor of 0.5 for
assumed dust temperatures that range from $10$ to $20$ K (Shirley et al. 2000).  
The variation is less severe when two wavelengths longer than $1$ mm are compared;
nevertheless, a suitable single dust temperature must be found. 

In reality there are strong gradients
in the density and temperature increasing toward the center of the core.
Since the dust continuum emission
of the envelopes of four of the sources in this survey have been modeled using
radiative transfer, we may use the calculated temperature profiles
$T(s)$ and constrained density profiles $n(s)$ along each line-of-sight distance, $s$,
to estimate the appropriate
characteristic dust temperature within an aperture.
We define the isothermal envelope temperature, $T_{iso}^{env}$, 
as the single dust temperature that characterizes the observed emission from
density and temperature gradients within a central aperture.  
For a telescope with normalized beam pattern, $P_n(\theta, \phi)$,
the isothermal envelope temperature in a central aperture is derived from the 
equation for specific intensity of optically thin dust emission convolved
with the telescope beam pattern,
\begin{equation}
B_{\nu}(T_{iso}^{env}) \int_{\Omega} \int_{s} P_n(\theta,\phi) n(s) ds d\Omega  = 
\int_{\Omega} \int_{s} P_n(\theta,\phi) B_{\nu}[T(s)] n(s) ds d\Omega \;\;,
\end{equation}
where $B_{\nu}$ is the Planck function.  Solving for $T_{iso}^{env}$ gives,
\begin{equation}
T_{iso}^{env} = (h\nu/k) \left[ \ln \left(1 + \frac{\int_{\Omega} P_n(\theta,\phi) N(\theta,\phi) d\Omega}{\int_{\Omega} P_n(\theta,\phi) \int_{s} \frac{n(s) ds}{exp(h\nu/kT(s)) - 1} d\Omega }\right) \right]^{-1} \;\;,
\end{equation}
where $N(\theta,\phi)$ is the column density at an impact parameter $\theta$ away from
the protostar.
The line-of-sight distance is related to the
impact parameter $\theta$ geometrically by $s^2 + \theta^2 = r^2$ where $r$ is the
radial distance from the protostar (see Adams 1991, Shirley et al. 2003).

The isothermal envelope temperatures for the best-fit one-dimensional dust
continuum models of Class 0 sources in this survey are shown in Figure 2.
\tie\ depends on the frequency of the observations.  The \tie\ curves
in Figure 2 flatten at millimeter wavelengths.  As a result, it is a good
assumption to assume a single characteristic dust temperature at
both wavelengths in Equation 2 as long as both of those wavelengths are
greater than $0.6$ mm ($\Delta T_{iso}^{env} < 0.5$ K).  
The single temperature assumption starts to break down
at submillimeter wavelengths, although the variation in \tie\ is not strong.
For instance, the typical difference in \tie\ between \textit{Herschel Space
Observatory} SPIRE wavelengths ($250 - 500$ \micron ) is slightly less than 2 K.
If the emission within the central aperture is dominated by envelope emission,
then the curves in Figure 2 constrain the appropriate dust temperature to use
in calculating $\beta$. In \S3.2, we explore the effects of the contribution
of the disk. 

We constrain the dust opacity index using Equation 2 from a plot of $\beta$ versus
the characteristic dust temperature (Figure 3).  The $\beta$ curves for
the 0.86 to 3.3 mm flux ratio (blue curves) and 1.25 to 3.3 mm flux ratio (red curves) 
are shown as solid lines in Figure 3.  The dashed lines represent the $\pm 1 \sigma$ 
statistical uncertainty in the flux at each wavelength.  At the characteristic 
\tie\ $\sim 16$ K for this sample 
of  Class 0 protostars,  $\beta$ does not have a strong dependence on the
dust temperature.  In general, the derived opacity index agrees within the statistical
calibration uncertainty between $\beta_{0.86/3.3}$ and $\beta_{1.25/3.3}$.  Typical
values range from $\beta_{mm} = 0.8$ to $2.2$ with an average value of
$\mean{\beta_{mm}} = 1.5 \pm 0.4$.  In the next section, we
interpret the derived $\beta_{mm}$ by accounting for the contribution from disk and envelope
emission.

In all cases except for L1527, the $\beta_{0.86/3.3}$
curve agrees within the statistical errorbars with 
the $\beta_{1.25/3.3}$ curve.  For L1527, the offset may be accounted
for by a systematic
calibration error of $20$\% 
at one or more of the three wavelengths (e.g., with SCUBA, MAMBO, and 
or MUSTANG calibration).  
The dominant source of uncertainty in determining $\beta_{mm}$
is the uncertainty in the flux ratio at two wavelengths.  The uncertainty is lower
if the two wavelengths are more widely spaced (e.g. lower for the 
ratio $0.86/3.3$ vs. $1.25/3.3$).  An
uncertainty in the fluxes of $20$\% results in a typical uncertainty of $\pm 0.4$ in  
$\beta_{1.25/3.3}$ and $\pm 0.3$ in $\beta_{0.86/3.3}$.  An accurate flux calibration
at both wavelengths and deep photometry is required to minimize this uncertainty.
A second possibility for this discrepancy is that the emission between 0.86 mm
and 3.3 mm is mixing different fractions of grain populations from the envelope
and disk.  This possibility is explored in the next section.

\subsection{Estimating the Disk Contribution}

The total flux observed at 3.3 mm is the sum
of emission from the protostellar envelope, disk, and wind (or jet)
\begin{equation}
S_{\nu}^{dust} = S_{\nu}^{env} + S_{\nu}^{disk} + S_{\nu}^{ff} \;\;\;.
\end{equation}
At submillimeter wavelengths, the dust continuum emission from 
Class 0 sources are expected to be dominated by their massive envelopes.
At millimeter wavelengths, this assumption may no longer be valid and the
disk emission may be a significant fraction of the total emission in the 
central beam.  At centimeter wavelengths, thermal
radio continuum emission (free-free emission) from the protostellar
jet or wind becomes the dominant emission mechanism (Anglada 1995). 
The centimeter wavelength free-free emission can be variable from Class 0 
protostars (see Shirley et al. 2007).  Unfortunately, we do not
have simultaneous centimeter continuum observations at the same epoch as
the MUSTANG observations; however, extrapolation from published VLA
fluxes and spectral indices indicate that the expected contribution from 
free-free emission at 3.3 mm is expected to be small ($< 10$\%). 
Therefore, we ignore the free-free contribution to the 3.3mm fluxes
in the following analysis.

While it may be advantageous to study dust properties at
millimeter wavelengths close to the Rayleigh-Jeans limit, 
the interpretation of single-dish observations
becomes more difficult since
derived $\beta_{mm}$ are an amalgamation of disk plus envelope opacities.
This is a particular issue for calculating $\beta$ since dust grains
may undergo coagulation in the dense enviornments of Class 0 disks,
and therefore the resulting $\beta_{mm}$ is expected to be lower
than for dust in the protostellar envelope (e.g., Henning \& Stognienko 1996, 
Dominik \& Tielens 1997,
Poppe et al. 2000, Draine 2006, Birnstiel et al. 2010).  We can identify the disk
contribution at (sub)millimeter wavlengths from the visibility amplitudes
of interferometric observations.  Since disks are typically small
($R < 100$ AU corresponding to $\theta < 1$\as ), they appear as
unresolved structures on baselines shorter than the characteristic
size.  The disk flux may be determined from the flux level of
a flattening in the visibility amplitudes, and a spectral index from
interferometric observations at two wavelengths.  In this section,
we estimate the fraction of disk emission at $3.3$ mm 
and the
impact on our interpretation of $\beta_{mm}$
for four of
the sources (L483, L1527, B335, and L1448C)
with published multi-wavelength interferometric observations.


L483 is perhaps the easiest example to analyze as the emission from
a disk is thought to be very weak for this source,
even at wavelengths as long as 3 mm (J\o rgensen et al. 2004, 2007, 2009).
This source has been observed with the SMA and OVRO at wavelengths
ranging from 0.8 mm to 3.0 mm and no evidence for a compact component
is seen in the visibility amplitudes.  J\o rgensen et al. (2009) estimates a negligible
disk mass compared to the envelope mass.  
The measured spectral index for L483 observed in this
paper is consistent
with the spectral index observed by J\o rgensen et al. (2007) with the SMA
between 0.8 and 1.25 mm ($\alpha_{0.8/1.25} = 3.7$ on baselines $> 40$ k$\lambda$).  
Therefore, our observations at 3.3 mm are probing the envelope structure
and are not significantly contaminated by emission from a disk.
The observed range of $\beta_{mm} = 1.57 - 2.00$ is consistent with the
opacities typically assumed
for radiative transfer models of Class 0 envelopes.  For instance,
the widely used OH5 opacities have a $\beta_{OH5} = 1.85$ for
coagulated dust grains at a density
of $10^6$ cm$^{-3}$ for $10^5$ years with thin ice mantles (Table 2,
column 5 of Ossenkopf \& Henning 1994; also see Table 2 of Shirley et al. 2005
for a summary of $\beta$ for various dust models).

In contrast, the emission from L1527 at 3.3 mm within a central MUSTANG
aperture appears to be dominated by the
disk emission.  L1527 has also been observed by J\o rgensen et al (2007) 
with the SMA at 0.8 and 1.3 mm where a very flat spectral index of 
$\alpha_{0.8/1.3} = 1.9$ is observed on baselines $> 40$ k$\lambda$.
Observations of this source have also been made using BIMA with the combination
of four array
configurations at 2.7 mm where a distinct flattening in visibility amplitudes
is observed on baselines $> 10$ k$\lambda$ with a flux of $42$ mJy 
(Y. Shirley, unpublished observations).  
The observed spectral index between 1.3 and 2.7 mm is consistent with 
SMA results ($\alpha_{1.3/2.7} = 1.8$). All of the observed emission in the MUSTANG
central beam may be accounted for by extrapolating this disk flux and 
interferometric spectral indices to 3.3 mm.  The spectral indicies reported in Table 1
are larger than the interferometric spectral indices because a significant fraction
($\geq 50$\%) of the flux in the single dish apertures at 0.86 and 1.25 mm is still 
coming from the envelope.  If the envelope dust has a steeper opacity index
than the disk dust opacity index, the resulting $\beta_{mm} = 0.82 - 1.42$ is then an 
overestimate of the true $\beta$ in the disk.  Despite this uncertainty in the true disk $\beta$, 
the low value is consistent with evidence of evolution of the dust properties
in the L1527 disk and indicates that different dust opacities than are used to model
the envelope are needed (cf. Shirley et al. 2002).

B335 is a popular target with extensive interferometric observations;
however, not all of these observations agree on the disk contribution.
B335 was studied extensively at 1.2 and 3.0 mm 
with the Plateau de Bure Interferometer (PdBI) 
by Harvey et al. (2003).  They found a slight flattening in the visibility 
amplitudes for baselines greater than $60$ k$\lambda$.  The observed flux on these 
long baselines was $21 \pm 3$ mJy at 1.2 mm and $\approx 2$ mJy at 3.0 mm.  The
resulting spectral index is $2.6$, slightly less than the spectral index of $3.1$ 
observed in this paper.  Extrapolation of the 3.0 mm disk flux to 3.3 mm results in a
small contribution to the MUSTANG flux ($1.7$ mJy or $10$\% of the 3.3 mm flux).  
The PdBI observations do not
agree with the flux estimates from extrapolation of the J\o rgensen et al. (2007) SMA
observations on baselines greater than $40$ k$\lambda$ which predict
a $3.3$ mm disk flux of $7$ mJy or approximately $50$\% of the MUSTANG flux.
The real contribution of the disk flux for B335 is probably somewhere in between
these estimates.  It is likely that the J\o rgensen et al. (2007) fluxes
include a contribution from the envelope since emission has been seen on 
baselines longer than $40$ k$\lambda$ from the envelope toward B335 (Harvey
et al. 2003). 

With the caveat that our estimates of $\beta_{mm}$ toward B335 may have some contribution ($10 - 50$\%)
from a disk component, the derived opacity index $\beta_{mm} = 1.18 - 1.45$ 
is significantly lower than has been
observed in the outer envelope (Shirley et al. 2010).  A comparison
of the opacity ratios at $442/2.2$ \micron\ and $862/2.2$ \micron\
from near-infrared extinction observations and submillimeter continuum images
yields a $\beta_{submm} = 2.1 - 2.5$ for lines-of-sight greater than $15$\as\
from the protostar (lines-of-sight where background stars were detected in NICMOS observations
by Harvey et al. 2001).  
This may be direct evidence of a change in the opacity
in the inner envelope and disk of B335.  Our derived $\beta_{mm}$ is
consistent with the slope of OH2 dust ($\beta_{OH2} = 1.35$; Ossenkopf \& Henning 1994; Shirley
et al. 2005)
for coagulated dust grains with no ice mantles as might be expected in the warm
inner envelope of a Class 0 protostar. However, contribution from disk emission may be
responsible, in part, for this percieved lowering of $\beta_{mm}$.  If we subtract the maximal
disk flux from $1.25$ mm and $3.3$ mm photometry, then $\beta_{mm} = 1.6 - 2.4$
which is consistent with the Shirley et al. 2010 outer envelope determination.
Unfortunately, until observations are performed with an interferometer 
at two wavelengths that match the wavelengths of single dish continuum 
observations (e.g., 0.86 mm and 3.3 mm with ALMA), then this 
level of uncertainty in determining the envelope $\beta_{mm}$ in the 
central aperture toward B335 will persist.

Observations of L1448C were also made by J\o rgensen et al. (2007) with the
SMA.  \textit{Spitzer Space Telescope} imaging of this region (J\o rgensen et al. 2006) 
have discovered that this source is actually two
Class 0 sources in $8$\as\ proximity (L1448C(N) and L1448C(S)).  
Because the southern source is
significantly weaker than the northern source (only $7$\% of the flux of the northern
source in the 1.25 mm SMA observations; J\o rgensen et al. 2007), it is likely that
the MUSTANG fluxes are predominantly from the northern source.
The observed spectral index between 0.8 and 1.25 mm on baselines greater
than $40$ k$\lambda$ is significantly shallower than the spectral indices
observed with single dish telescopes.  Extrapolating the SMA results to
$3.3$ mm indicates that as much as $50$\% of the MUSTANG flux may be
due to the emission from the disk.
Again, without multi-wavelength interferometric observations that
match the wavelength of single-dish observations, we cannot accurately
determine
the envelope $\beta_{mm}$.  The range of $\beta_{mm} = 1.10 - 1.52$ 
observed toward L1448C is very similar to the range observed toward B335.  
Accounting for the maximal disk contribution can also increase the
range of $\beta_{mm}$ in the envelope by $0.5$.

Unfortunately, there are no published u,v amplitude curves
for L1448N and L1448NW; therefore, we are unable
to assess the disk contribution to the MUSTANG fluxes in these two cases.
Interferometric observations have been made at 1.3 and 2.7mm toward
the L1448 IRS3 region which includes these two sources (Looney et al. 2000,
Kwon et al. 2006).  L1448N is comprised of two bright sources
separated by 10\as\ (labeled L1448 IRS3 A
and B in Looney et al. 2000) which makes separating their visibility
amplitudes very difficult.
We note that L1448NW has the highest range in $\beta_{mm} = 1.94 - 2.23$.
Given such high values and the expectation that the opacity index is lower
in disks, it seems unlikely that L1448NW has a significant disk contribution.
However, the only way to confidently constrain the envelope and disk
opacities is to analyze the visibility amplitudes at multiple wavelengths with an 
interferometer.

\section{Summary}

We have observed 6 Class 0 protostars with the MUSTANG camera at 3.3 mm.
We report fluxes and calculate the spectral indicies 
at millimeter wavelengths.  Systematic flux uncertainties of up to $20$\%
at (sub)millimeter wavelengths limit determinations of $\beta_{mm}$ to
$\pm 0.4$ if an appropriate characteristic dust temperature is used.
We have estimated the characteristic isothermal temperature
in a central beam ($11$\as ) from previously published dust continuum 
radiaitve transfer models ($\mean{\tie } = 16$ K).  
The disk emission fraction at $3.3$ mm was estimated 
from published interferometric observations. We found emission
at $3.3$ mm is dominated by the envelope for L483 and by the disk
for L1527.  The envelope $\beta_{mm}$
for L483 is between $1.6$ to $2.0$, consistent with the opacity index
for the widely used OH5 dust opacities for Class 0 envelopes
($\beta_{OH5} = 1.85$).  
The disk $\beta_{mm} \leq 0.8$ to $1.4$ 
for L1527 is flatter than typical envelope opacity indicies likely
indicating that grain growth is occuring in the disk of L1527.
B335 and L1448C may have comparable disk and envelope
emission although interferometric observations are needed to
better constrain the emission fraction.  Taking the maximal disk
contribution at 3.3 mm into account for B335 leads to an estimate
of the envelope $\beta_{mm} = 1.6 - 2.4$ that is consistent with the
recent $\beta_{submm}$ 
determination in the outer envelope by Shirley et al. (2010).

The 3.3mm obsevations in this paper should be used in conjunction with future
interferometric observations to constrain the properties of the dust emission
in the envelope and disk of Class 0 sources.  The next step required
in understanding the physical structure of these sources is muti-dimensional
dust continuum radiative transfer which includes the envelope, disk, and
outflow cavity with dust properties constrained from multi-wavelength
interferometric and single-dish observations.  With the incredible sensitivity
of ALMA over a wide range of baselines, it will be possible to obtain
the necessary observations at wavelengths that are well matched to
current single-dish bolometer cameras.  The MUSTANG observations presented
in this paper have extended this wavelength
coverage well into the millimeter.

\acknowledgements{Acknowledgements}
We graciously thank the UA/NASA Space Grant Undergraduate 
Research Internship Program for funding
David E. Bolin.  The authors would like to thank the MUSTANG instrument team from 
the University of Pennsylvania, Cardiff University, NASA-GSFC, NRAO and NIST for 
their efforts on the instrument and software that have made this work possible.
We thank the operators of the Green Bank Telescope for their assistance
during the observations.  We are especially grateful to Todd Hunter whom was present
during the February 2009 observations to test the OOF holography and greatly increase the
GBT aperture efficiency for our MUSTANG observing.  We also thank the referee for
a speedy response and many useful comments that benefited this paper.




\begin{deluxetable}{llrclcccccc}
\rotate
\tiny
\tablecolumns{11}
\tabletypesize{\footnotesize}
\tablecaption{Class 0 Millimeter Photometry\tablenotemark{a}\label{tab1}}
\tablewidth{0pt} 
\tablehead{
\colhead{Source}            & 
\colhead{$\alpha$ (J2000.0)} &
\colhead{$\delta$ (J2000.0)} &
\colhead{$S_{0.86 mm}$\tablenotemark{b}}      &
\colhead{$S_{1.25 mm}^{peak}$\tablenotemark{c}}      &
\colhead{$S_{3.3 mm}^{peak}$}      &
\colhead{$S_{3.3 mm}$}      &
\colhead{$\alpha_{0.86/3.3}$}	    &
\colhead{$\alpha_{1.25/3.3}$}	    &
\colhead{$T_{iso}^{env}$ \tablenotemark{d}}       &    
\colhead{$\beta_{mm}^{dust}$ \tablenotemark{e}}  \\
\colhead{} &
\colhead{($^h$~~$^m$~~$^s$~)~} &
\colhead{($\degree$ ~\arcmin\ ~\arcsec)} &
\colhead{(mJy 20\as )} &
\colhead{(mJy/beam )}  &
\colhead{(mJy/beam )}  &
\colhead{(mJy 20\as )} &
\colhead{} &
\colhead{} &
\colhead{(K)} &
\colhead{}    
}
\startdata 
L1448NW & 03 25 35.8 & $+$30 45 34 & 2510 (160) & 560 (25) & 13.5 (1.4) 	& ...		& ...		& 3.83 (0.11)		& ...	& 1.94 - 2.23		\\
L1448N  & 03 25 36.3 & $+$30 45 15 & 4610 (290) & 1400 (5) & 85.8 (8.6) 	& 100.5 (10.1)	& 2.85 (0.12)	& 2.88 (0.11)		& ...	& 1.02 - 1.29		\\
L1448C  & 03 25 38.8 & $+$30 44 03 & 1970 (130) & 620 (15) & 29.1 (2.9) 	& 39.2 (3.9) 	& 2.92 (0.12)	& 3.15 (0.13)		& 16.7	& 1.10 - 1.52	\\
L1527   & 04 39 53.9 & $+$26 03 10 & 1690 (110) & 375 (6)  & 28.7 (2.9)	 	& 33.3 (3.4)	& 2.92 (0.12)	& 2.65 (0.12)		& 14.1	& 0.82 - 1.42	\\
L483 	& 18 17 29.8 & $-$04 39 38 & 1870 (100) & 290 (15) & 8.7 (0.9) 		& 19.3	(1.9)	& 3.41 (0.12) 	& 3.61 (0.16)		& 18.1	& 1.57 - 2.00	\\
B335 	& 19 37 01.1 & $+$07 34 11 & 1400 (80)  & 270 (5)  & 14.4 (1.5) 	& 25.8 (2.6)	& 2.98 (0.12) 	& 3.02 (0.16)		& 14.4	& 1.18 - 1.45	\\
\enddata
\tablenotetext{a}{The FWHM beam sizes at each wavelength are 16\as\ at 0.86 mm, 
11\as\ at 1.25mm, 11\as\ at 3.3mm.  The notation (mJy $20$\as )
means the flux density (mJy) observed in a 20\as\ diameter aperture. The positions are the $862$ \micron\
continuum peak positions from Shirley et al. (2000).}
\tablenotetext{b}{Aperture photometry determined from Shirley et al. (2000) 862 \micron\ images.  Statisitcal errorbars are tabulated.}
\tablenotetext{c}{Photometry reported in M\"otte \& Andr\'e (2001).}
\tablenotetext{d}{The characteristic isothermal temperature (Equation 4) in a $11$\as\
aperture between 0.86 and 3.3 mm (see Figure 2).}
\tablenotetext{e}{The range in $\beta_{mm}$ determined between 0.86, 1.25, and 3.3 mm at \tie\ (see Figure 3). \tie $= 16$ K was assumed for L1448N and L1448NW.}
\end{deluxetable}

\clearpage

\begin{figure}
   \centering
   \vspace*{5in}
   \leavevmode
   \includegraphics{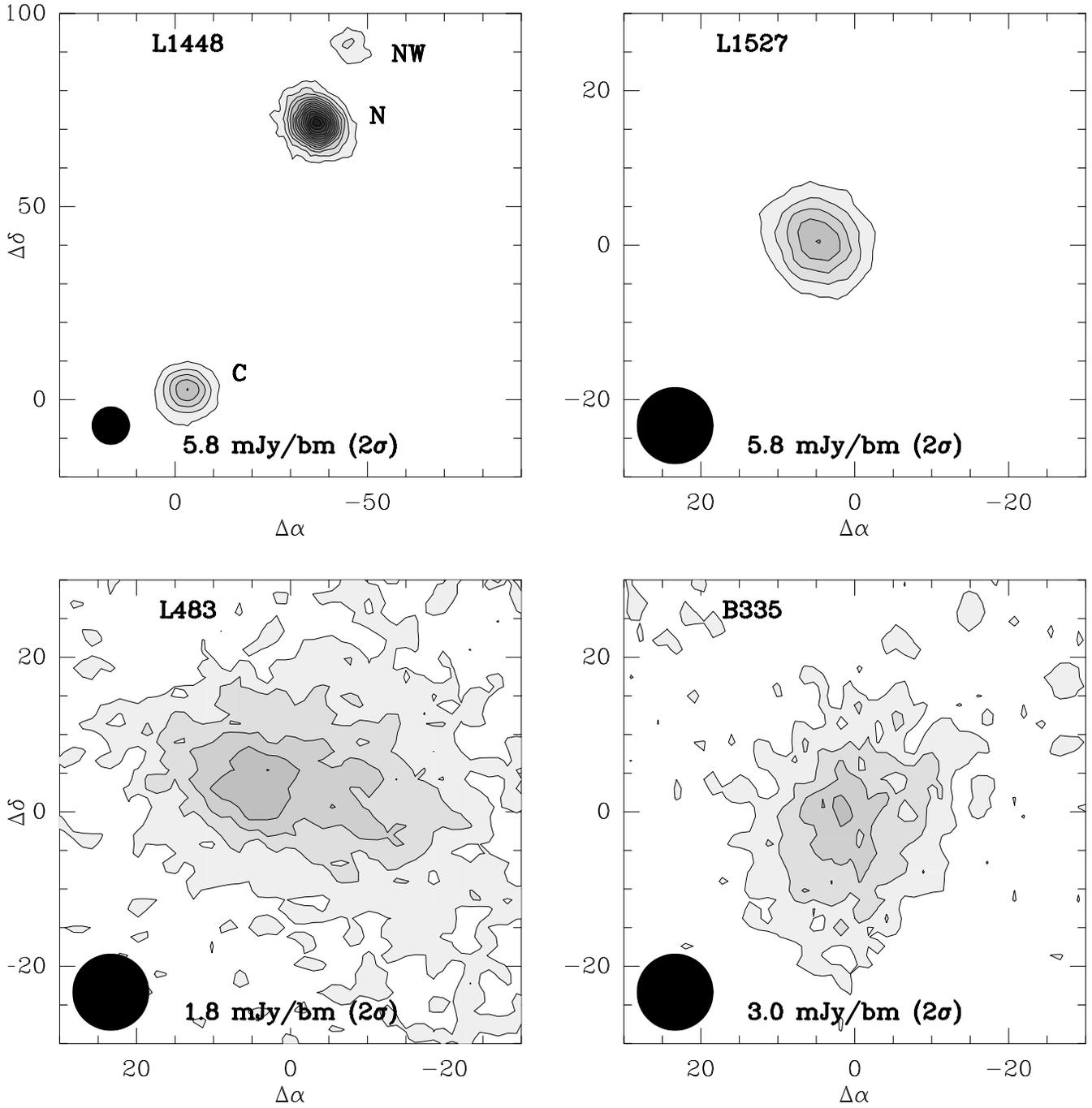}
 \vskip 2.0in
\caption{MUSTANG 3.3 mm contour maps of the six Class 0 protostars observed.  Clockwise from 
the top left: L1448C, L1448N, and L1448NW (upper left panel); L1527; L483; and B335.  The contours are $2 \sigma$ and are listed at the bottom of 
each panel.  The central coordinates (0,0) are listed in Table 1.}
\end{figure}

\begin{figure}
   \centering
   \vspace*{5in}
   \leavevmode
   \includegraphics{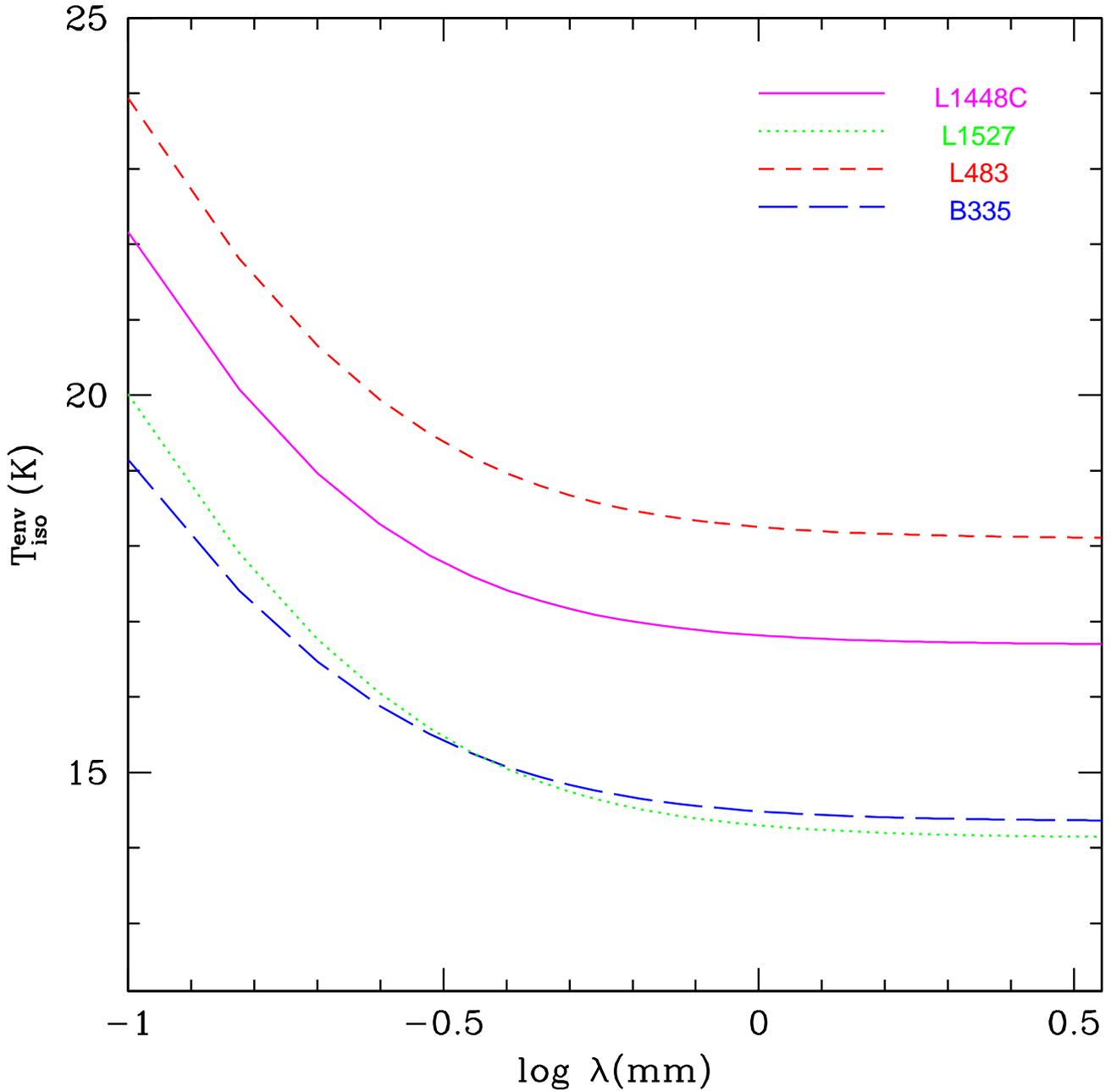}
 \vskip 2.0in
\caption{The isothermal characteristic envelope dust temperature in an $11$\as\ 
beam for Class 0 sources.  The curve for each source represents the best-fitted dust continuum radiative transfer models from Shirley et al. (2002, 2010). Notice that the curves flatten at millimeter 
wavelengths.}
\end{figure}

\begin{figure}
   \centering
   \vspace*{5in}
   \leavevmode
   \includegraphics{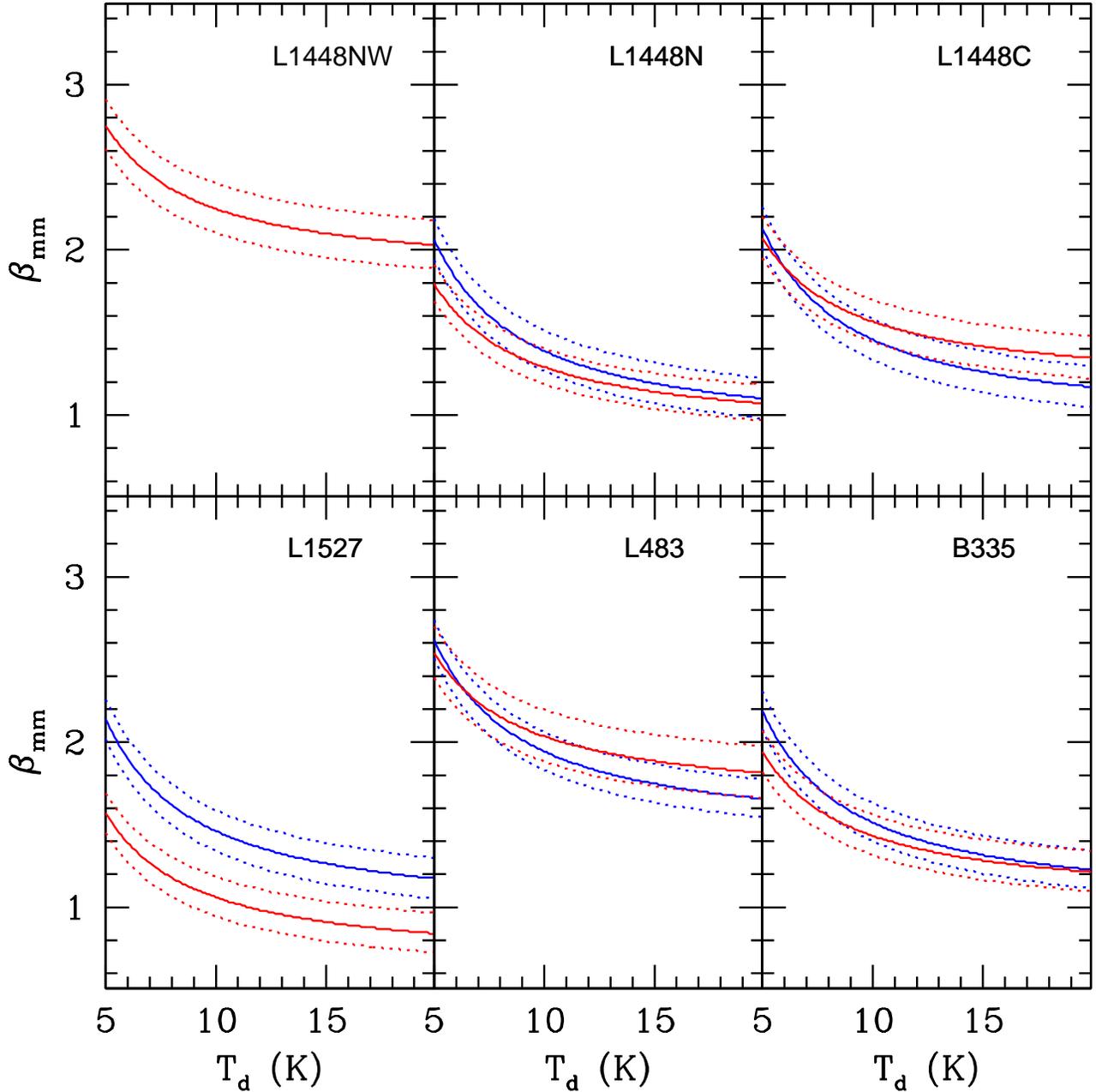}
 \vskip 2.0in
\caption{$\beta_{mm}$ plotted versus the characteristic dust temperature within an aperture.
$\beta_{mm}$ is shown for the 1.25/3.3 mm flux ratio (red curves) and 0.86/3.3 mm flux ration
(blue curves).  The dashed lines represent $\pm 1\sigma$ uncertainties in the flux ratios.}
\end{figure}

\end{document}